%% file: cggw.tex
\documentclass[%
 preprint,amsmath,amssymb,aps,superscriptaddress,prl]{revtex4-2}

\usepackage{graphicx}
\usepackage{bm}
\usepackage{subfig}
\usepackage{float}
\usepackage{caption}

\usepackage{color}
\usepackage[normalem]{ulem}

\begin{document}

\title{Linear Response Functions Respecting Ward-Takahashi Identity and Fluctuation-Dissipation Theorem within $GW$ Approximation}

\author{Hui Li}
\affiliation{%
School of Physics, Peking University, Beijing 100871, China
}
\author{Zhipeng Sun}%
 \email{zpsun@csrc.ac.cn}
\affiliation{
Beijing Computational Science Research Center, Beijing 100193, China
}%

\author{Yingze Su}%
\affiliation{%
School of Physics, Peking University, Beijing 100871, China
}

\author{Haiqing Lin}%
 \email{haiqing0@csrc.ac.cn}
\affiliation{
Beijing Computational Science Research Center, Beijing 100193, China
}

\author{Huaqing Huang}%
 \email{huanghq07@pku.edu.cn}
\affiliation{%
School of Physics, Peking University, Beijing 100871, China
}

\author{Dingping Li}%
 \email{lidp@pku.edu.cn}
\affiliation{%
School of Physics, Peking University, Beijing 100871, China
}

\date{\today}

\begin{abstract}
The calculation of response functions in correlated electronic systems is one of the most important problems in the condensed matter physics. To obtain a physical response function, preserving both the Ward-Takahashi identity and the fluctuation-dissipation theorem are crucial. Here we propose a self-consistent many body method within the GW framework to calculate the response functions based on the fluctuation-dissipation theorem, which also satisfies the Ward-Takahashi identity.
The validity of this methodology is demonstrated on the two-dimensional one-band Hubbard model, where both the Ward-Takahashi identity and fluctuation-dissipation theorem are verified numerically. Moreover, comparing to the accurate spin susceptibility of the determinantal Monte Carlo approach, the results obtained from our method are quite satisfactory and the computational cost are greatly reduced.
\end{abstract}

\maketitle
\textit{Introduction.}---The calculation of the correlation functions is a long-standing problem in condensed matter physics; they are related to the transport properties of materials and can be measured in experiments \cite{coleman_2015}. The exact solution of the correlation functions does respect the Ward-Takahashi identities presenting for the symmetries\cite{book-peskin}. However, most approximations do not guarantee this constraint and thus lose the crucial information on the symmetries of the systems. Such calculations may lead to quantitatively inaccurate results and even qualitatively unphysical solutions \cite{Parcollet2014,Kadanoff}. Various schemes are proposed to preserve the Ward-Takahashi identities(for example in \cite{Levin2014,Levin2018,Millis2009,Mravlje2019}), however, it is also crucial to respect the fluctuation-dissipation theorem (FDT), which relates the response function to the correlation functions\cite{Kubo_1966}. The deviation of the FDT will result in contradictions \cite{FLEXFDT}, and thus lead to unphysical solutions.

A physical correlation should be defined as the response of a physical quantity to an external source originally due to Ornstein and Zernike\cite{Orns14}. This basic physical doctrine was applied to the mean-field analysis in the Ising model in classical statistics, where the physical spin susceptibility is calculated by the functional derivative of the mean local spin with respect to the external magnetic field (for example in \cite{Peliti2011}). In this scheme, the FDT is satisfied by definition, and the conservation laws are preserved automatically. In quantum field theory, the covariant Gaussian approximation is also based on a similar doctrine, where the propagator is obtained by the functional derivative of the vacuum expectation value with respect to the external source. As a result, this method preserves the extremely important Goldstone theorem and all higher-order WTIs\cite{PhysRevD.39.2332,PhysRevD.40.504,WANG2017228,PhysRevB.104.125137}. Therefore, in order to preserve the FDT and the WTI, the physical correlation functions should be calculated by the functional derivatives with respect to the relevant external source.

In this Letter, we apply the functional derivative scheme to the generalized $GW$ approximation (GGW)\cite{Hedin,RevModPhys.74.601,Aryasetiawan_1998,Aryasetiawan2008,Aryasetiawan_2009,Yuchen2016,Lucia2018,Patrick2019}, which was developed for the systems including the spin-dependent interaction. This approach can preserve the WTI and FDT for general cases, which are also verified numerically in the repulsive two-dimensional Hubbard model. The computational complexity is of the same order of the widely used Bethe-Salpeter equations (BSE) within the GW method, and thus is expected to be applicable for a wide range of realistic material calculations.

\textit{GGW approximation in general cases.}
---The generalized $GW$ approximation was proposed to deal with explicitly spin-dependent interaction, and can be applied to various kinds of electronic systems. We reformulate it in the functional path integral formalism, and start with the Matsubara action:
\begin{eqnarray}
    S[\psi^*,\psi]=&-&\sum_{\alpha_1\alpha_2}\int d(12)\psi_{\alpha_1}^*(1)T_{\alpha_1\alpha_2}(1,2)\psi_{\alpha_2}(2)\nonumber\\
    &+&\frac{1}{2}\int d(12)\sum_{ab}\sigma^a(1)V^{ab}(1,2)\sigma^b(2).\label{S}
\end{eqnarray}

Here, the charge/spin composite operator $\sigma^a(1)=\sum_{\alpha\alpha'}\psi^*_{\alpha}(1)\tau^a_{\alpha\alpha'}\psi_{\alpha'}(1)$, $\tau^a(a=0,x,y,z)$ are Pauli matrices, Greek letters like $\alpha$ indicate spin up and spin down. $\psi,\psi^*$ are  Grassmannian fields.
Notation $(1) = (\tau_1,\vec x_1)$ contains the space coordinate $x_1$, and the imaginary time coordinate $0\leq\tau_1\leq \beta$, where $\beta$ is the inverse temperature. The integral over $(1)$ stands for integral or sum over all space and time coordinates. The two-body interaction is symmetric, i.e., $V^{ab}(1,2)=V^{ba}(2,1)$, and it can describe the usual Coulomb interaction, the spin-spin interaction, and the spin-orbit interaction.

The one-body Green's function is defined in an ensemble average form:
\begin{equation}
    G_{\alpha_1\alpha_2}(1,2) = \left\langle \psi^*_{\alpha_2}(2)\psi_{\alpha_1}(1) \right\rangle.
\end{equation}
Here $\left< \cdots \right>$ presents for $\frac{1}{Z}\int D[\psi^*,\psi]\cdots e^{-S}$, with $Z=\int D[\psi^*,\psi] e^{-S}$ the grand partition function. Since the interaction has a spin structure, it is convenient to denote the matrix in the spin space as:
\begin{equation}
    \underline{X}=
    \begin{bmatrix}
    X_{\uparrow\uparrow}& X_{\uparrow\downarrow}\\
    X_{\downarrow\uparrow}&  X_{\downarrow\downarrow}
    \end{bmatrix}.
\end{equation}
Note that its trace is denoted by $\mathrm{Tr}[\underline{X}]= X_{\uparrow\uparrow}+X_{\downarrow\downarrow}$.

Then one can derive the generalized Hedin's equations for the action Eq.~(\ref{S}), and the lowest approximation for the Hedin's vertex function leads to the GGW approximation (for details, see Supplemental Material \footnote{\label{fn}See Supplemental Material, for more details about the derivation, the proof of the WTI and the algorithm.}).
The full Green's function, $\underline{G}$, is determined from the bare Green’s function $\underline{T}$ and self-energy, through Dyson’s equation: $\underline{G}^{-1}(1,2) = \underline{T}(1,2)-\underline{\Sigma}_\mathrm{H}-\underline{\Sigma}(1,2)$. In the GGW approximation, the Hartree self-energy is given by
\begin{equation}
    \underline{\Sigma}_\mathrm{H}(1,2)=\delta(1,2)\sum_{ab}\int d(4)\underline{\tau}^aV^{ab}(1,4)\mathrm{Tr}[\underline{\tau}^b \underline{G}(4,4)],
\end{equation}
and the GGW self-energy is given by
\begin{equation}
    \underline{\Sigma}(1,2) =-\sum_{ab}\underline{\tau}^a \underline{G}(1,2)\underline{\tau}^bW^{ba}(2,1),
\end{equation}
where $W$ is the dynamic effective charge/spin potential and determined by the polarization function $P$ through the relation  $(W^{-1})^{ab}(1,2) = (V^{-1})^{ab}(1,2)-P^{ab}(1,2)$. The polarization function is approximated by:
\begin{equation}
        P^{ab}(1,2)=\mathrm{Tr}[\underline{\tau}^a\underline{G}(1,2)\underline{\tau}^b\underline{G}(2,1)].
\end{equation}
These equations can be solved self-consistently. It is worth mentioning that, in the GGW approximation, the Green's function $\underline{G}$ and the self-energy $\underline\Sigma$ are spin-dependent, the screened potential $W$ and the polarization function $P$ are $4\times 4$ matrices containing the coherence between charge and spin channels. Next, we will address the problem of the two-body correlation functions.

\textit{Covariant scheme GGW method.}--- According to FDT, the two-body correlation functions should be defined as the response of the physical quantity in the presence of an external potential, which we refer as the covariant scheme. The scheme for calculating a general two-body correlation function $\chi_{XY}(1,2)=\left< X(1)Y(2)\right>$ within the GGW framework, where $X, Y$ are binary composite operators, is formulated as follows.

First, one adds the corresponding source term to the action, $S[\psi^*,\psi;\phi] = S[\psi^*,\psi]-\int d(1)\phi(1) X(1)$ and the correlation can be obtained by $\chi_{XY}(1,2)=\frac{\delta \left< Y(2) \right>}{\delta \phi(1)}$. Then, write down the off-shell GGW equations (keep $\phi\neq 0$), and calculate the functional derivative of the GGW equations with respect to $\phi$. Finally, let the source $\phi$ tend to $0$ to obtain the on-shell results. Although we restrict our discussion to the GGW, this scheme can also be applied to to different many body approaches

\begin{figure*}
\includegraphics{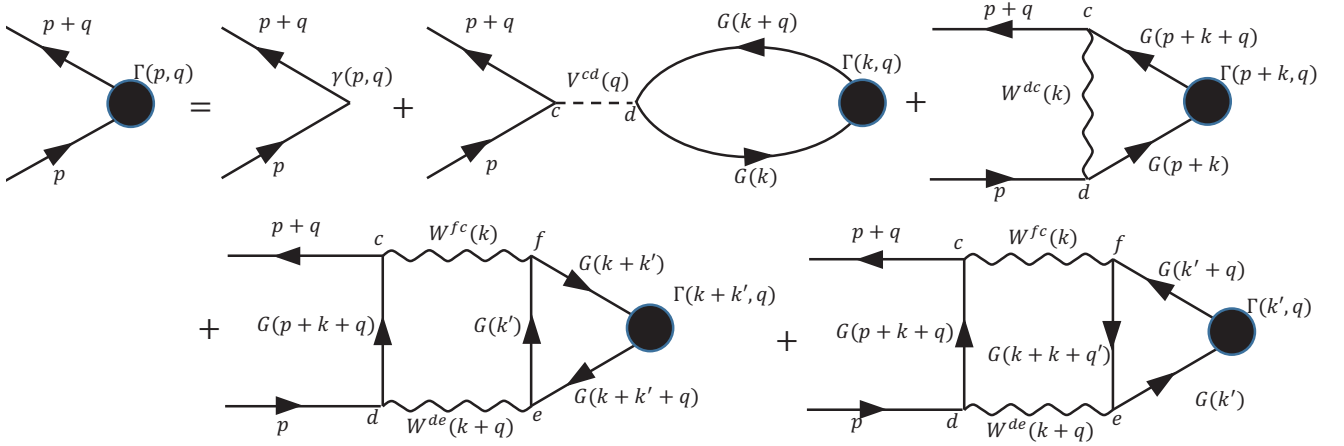}
\captionsetup{font={small},justification=raggedright}
\caption{\label{FDvertex} The Feynman diagram of the full cGGW vertex function in Eq.~(\ref{vertex}) for translation invariant systems in the momentum space.}
\end{figure*}

The functional derivative of the off-shell GGW equations with respect to the external source $\phi$ leads to the covariant GGW (cGGW) equations. The equation involves the full vertex function $\underline{\Gamma}_{\phi }(1,2,3)=\frac{\delta \underline{G}^{-1}(1,2)}{\delta\phi(3)}$, which consists of 5 terms shown in Fig.~\ref{FDvertex}:
\begin{eqnarray}
\underline{\Gamma}_{\phi}(1,2,3)=&&\underline{\gamma}_\phi(1,2,3)+\underline{\Gamma}_\mathrm{H}(1,2,3)+\underline{\Gamma}_\mathrm{MT}(1,2,3)\nonumber\\
&&+\underline{\Gamma}_\mathrm{AL1}(1,2,3)+\underline{\Gamma}_\mathrm{AL2}(1,2,3)\label{vertex}.
\end{eqnarray}
Here, the bare vertex $\underline{\gamma}_\phi$ depends on the operator $X$. In the charge/spin response case, for example,  $X^a=\sigma^a$, the bare vertex takes the form $\underline{\gamma}_\phi(1,2,3)=\underline{\tau^a}\delta(1,2)\delta(1,3)$. The ``bubble" vertex is induced by the Hartree self-energy and takes the form:
\begin{widetext}
\begin{equation}
    \underline{\Gamma}_\mathrm{H}(1,2,3) = -\delta(1,2)\sum_{cd}\int d(456)\tau^cV^{cd}(1,4)\text{Tr}[\tau^d \underline{G}(1,5)\underline{\Gamma}_{\phi }(5,6,3)\underline{G}(6,2)].
\end{equation}
\end{widetext}
Note that the conventional random-phase-approximation-like (RPA) formula only consists of the first two terms in Eq.~(\ref{vertex}).
The Maki-Thompson-like (MT) vertex and two distinct Aslamazov-Larkin-like (AL) vertices \cite{Larkinbook} represent the vertex corrections beyond the RPA, which take the form:
\begin{widetext}
\begin{equation}
    \underline{\Gamma}_\mathrm{MT}(1,2,3)=-\sum_{cd}\int d(45)\underline{\tau}^c\underline{G}(1,4)\underline{\Gamma}_{\phi }(4,5,3)\underline{G}(5,2)\underline{\tau}^d W^{dc}(2,1),
\end{equation}
\begin{equation}
    \underline{\Gamma}_\mathrm{AL1}(1,2,3)=-\sum_{cdef}\int d(4567)\underline{\tau}^c \underline{G}(1,2)\underline{\tau}^dW^{de}(1,4)\text{Tr}[\underline{\tau}^e\underline{G}(4,6)\underline{\Gamma}_{\phi }(6,7,3)\underline{G}(7,5)\underline{\tau}^f \underline{G}(5,4)]W^{fc}(5,2),
\end{equation}
\begin{equation}
    \underline{\Gamma}_\mathrm{AL2}(1,2,3)=-\sum_{cdef}\int d(4567)\underline{\tau}^c \underline{G}(1,2)\underline{\tau}^dW^{de}(1,4)\text{Tr}[\underline{\tau}^e \underline{G}(4,5)\underline{\tau}^f\underline{G}(5,6)\underline{\Gamma}_{\phi }(6,7,3)\underline{G}(7,4)]W^{fc}(5,2).
\end{equation}
\end{widetext}
Since the average $\left<Y^b(2)\right>$ is a function of the Green's function $G$, the two-body correlation function $\chi_{X^aY^b}(1,2)=\left< X^a(1)Y^b(2)\right>$ can be obtained by the vertex $\underline{\Gamma}_\phi$.

Such response functions satisfy the FDT by definition, and we have also theoretically proven that the WTI is preserved in our covariant scheme (see Supplemental Material \footnotemark[\value{footnote}] for more details). It is worthwhile noting that by neglecting two AL vertices in Eq.~(\ref{vertex}), the present approach reduces to the Bethe-Salpeter equation (BSE) in the GW region, which preserves the WTI but violates the FDT.

\textit{Implementation in the Hubbard model.}
---We now apply the cGGW method to the 2D ($L\times L$) one-band Hubbard model and use the discrete imaginary time algorithm \cite{book-Negele} to solve the cGGW equations. In the discrete time method, the integral over time $\int d(\tau)$ is replaced by a summation over time slides $\sum_{l=0}^{M-1}\Delta\tau$, where $\Delta\tau=\frac{\beta}{M}$ and $M$ is the number of the time slices.
The Hubbard Hamiltonian with the spin-dependent interaction takes the form\cite{Schafer2021}
\begin{eqnarray}
    \mathcal{H}=&&-t\sum_{\left< \vec{x}_{1}\vec{x}_{2}\right>\alpha}\psi_{\vec{x}_{1}\alpha}^{\dagger}\psi_{\vec{x}_{2}\alpha}\nonumber\\
    &&-\frac{U}{6}\sum_{\vec{x}a}\sigma_{\vec{x}}^{a}\sigma_{\vec{x}}^{a}-\left(\mu-\frac{U}{2}\right)\sum_{\alpha\vec{x}}\psi_{\alpha\vec{x}}^{\dagger}\psi_{\alpha\vec{x}},
\end{eqnarray}
where $\hat\psi^\dagger_{\vec{x}_{1}\alpha}$ creates an electron with spin $\alpha$ at lattice site $\vec{x}_{1}$. $t$ is the (nearest-neighbor) hopping amplitude and all energies
are given in units of $t=1$ in this paper. $\left< ij\right>
$ denotes summation over nearest-neighbor lattice sites, $U$ is the on-site interaction and $\mu$ is the chemical potential.

The Hubbard model is a translation invariant lattice system with the global charge $U(1)$ symmetry, and the corresponding WTI in the momentum space takes the form:
 \begin{eqnarray}
 &&\mathrm{i}\sum_{\nu=x,y}\underline\Gamma^\nu(p,q)[1-e^{\mathrm{i}q^\nu}]-\mathrm{i}\omega_q\underline\Gamma^0(p,q)\nonumber\\
 &&=\underline G^{-1}(p)-\underline G^{-1}(p+q)\label{WTI}.
 \end{eqnarray}
Here $\Gamma^\nu,\Gamma^0$ are the current vertex and the charge vertex. The fermionic momentum and Masubara frequency are denoted by $p=((p_x,p_y),\omega_p=(2m_p+1)\pi T)$, and the bosonic momentum and Masubara frequency are denoted by $q=((q_x,q_y),\omega_q=2m_q\pi T)$.
For the current operator \cite{smit_2002, Rosenstein}
\begin{equation}
    J^\nu(1)=\mathrm{i}t\sum_{\alpha=\uparrow,\downarrow}[\psi^*_{\alpha}(1+\hat e^{\nu})\psi_{\alpha}(1)-\psi^*_{\alpha}(1)\psi_{\alpha}(1+\hat e^{\nu})],
\end{equation}
the corresponding bare current vertex takes the form
\begin{equation}
    \underline{\gamma}^\nu(1,2,3)=\underline{\tau}^0it\delta(1,3)[\delta(1,2+\hat e^\nu)-\delta(1,2-\hat e^\nu)]\label{barecc},
\end{equation}
where $\hat e^\nu$ is the lattice vector. The self-consistent equations for the full current vertex $\Gamma^\nu$ are obtained by substituting Eq.~(\ref{barecc}) to Eq.~(\ref{vertex}).

To check the FDT which relates the response function to the correlation function, we focus on the antiferromagnetic fluctuation in the Hubbard model. We calculate the static antiferromagnetic spin susceptibility and the spin-spin correlation function $\chi_{sp}$ at momentum $\vec Q=(\pi, \pi)$, which should satisfy the following equality according to the FDT\cite{altland_simons_2010},
\begin{equation}
    \left.\frac{\partial m}{\partial h}\right|_{h=0}=\chi_{\mathrm{sp}}(\vec Q ,i\omega_n=0),
\end{equation}
where $h$ is the staggered field, $m$ is the staggered magnetization. $\chi_{\mathrm{sp}}(p)=\left<\sigma^z(p)\sigma^z(-p)\right>$ is the spin-spin correlation, which is calculated through the vertex $\Gamma_{\phi}$ related to the spin $\sigma^z$ within the cGGW: $\chi_{\mathrm{sp}}(p)=-\text{Tr}[\underline{G}(p+q)\underline\Gamma_{\phi}(q,p)\underline G(q)\tau^z]$.

\begin{figure}[htbp]
    \includegraphics[width=1\columnwidth]{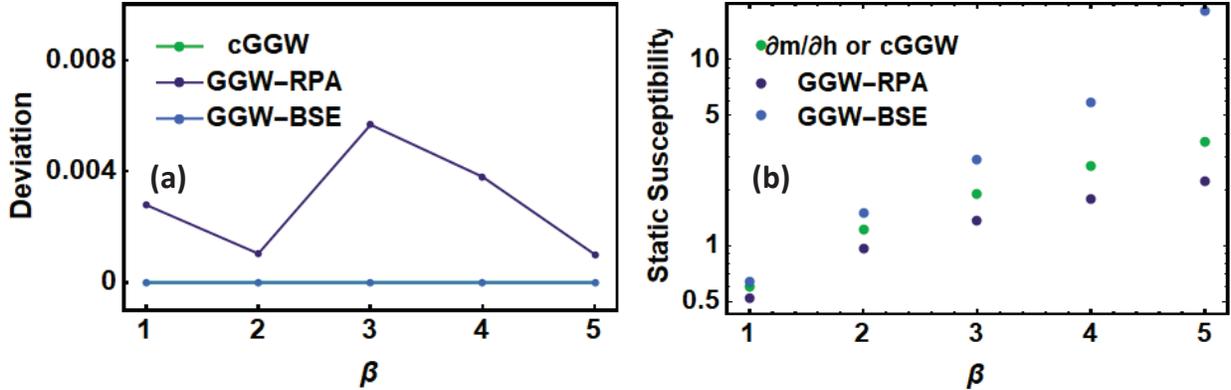}
    \captionsetup{font={small},justification=raggedright}
    \caption{\label{WandF} (a) shows the Deviation $D(p,q)$ of the WTI with momentum $\vec p = (\pi/2,\pi/2), \vec q=(\pi,\pi)$ and frequencies $m_p=1,m_q=2$ obtained by cGGW, GGW-RPA and GGW-BSE. The deviation for cGGW and GGW-BSE are 0, while GGW-RPA are not. (b) shows the comparison of the static antiferromagnetic spin susceptibility obtained by $\left.\frac{\partial m}{\partial h}\right|_{h=0}$, cGGW, GGW-RPA and GGW-BSE. Only the cGGW respects the FDT.}

\end{figure}
\textit{Numerical results.}--
By solving the cGGW equations of the Hubbard model, we can, in principle, obtain the two-body correlation functions for any values of the parameter set ($U, \beta, \mu, L$, and $M$). As a prototypical example, here we set a typical value of $U=2$, so as to compare with previous studies of the 2D Hubbard model using multiple methods \cite{PhysRevX.11.011058}.

To check the WTI and FDT, we calculate the half-filled Hubbard model on a $16\times 16$ cluster with $M=1024$. The deviation of the WTI is measured by
\begin{equation}
    D(p,q) = \left\Vert\frac{\mathrm{Tr}[LHS(p,q)-RHS(p,q)]}{\mathrm{Tr}[LHS(p,q)]}\right\Vert,
\end{equation}
where $(LHS)$ and $(RHS)$ are the left- and the right-hand sides in Eq.~(\ref{WTI}). As shown in Fig.~\ref{WandF}(a), the deviations $D(p,q)$ are negligible for cGGW and BSE vertex, verifying the WTI as expected. At the same time, RPA violates the WTI significantly, indicating a rather poor description of conversion laws.
For the FDT, we directly calculate the static antiferromagnetic susceptibility ($\partial m/\partial h$) by considering a staggered field $h$ in the GGW equations. Compared with the spin-spin correlation function $\chi_{\mathrm{sp}}$ obtained from different methods, it is found that only the cGGW method preserves the FDT at all temperature ranges, as shown in Fig.~\ref{WandF}(b). However, RPA and BSE lead to a violation of the FDT and, therefore, intrinsic inconsistencies in the calculation of response functions.

\begin{figure}
\includegraphics[width=1\columnwidth]{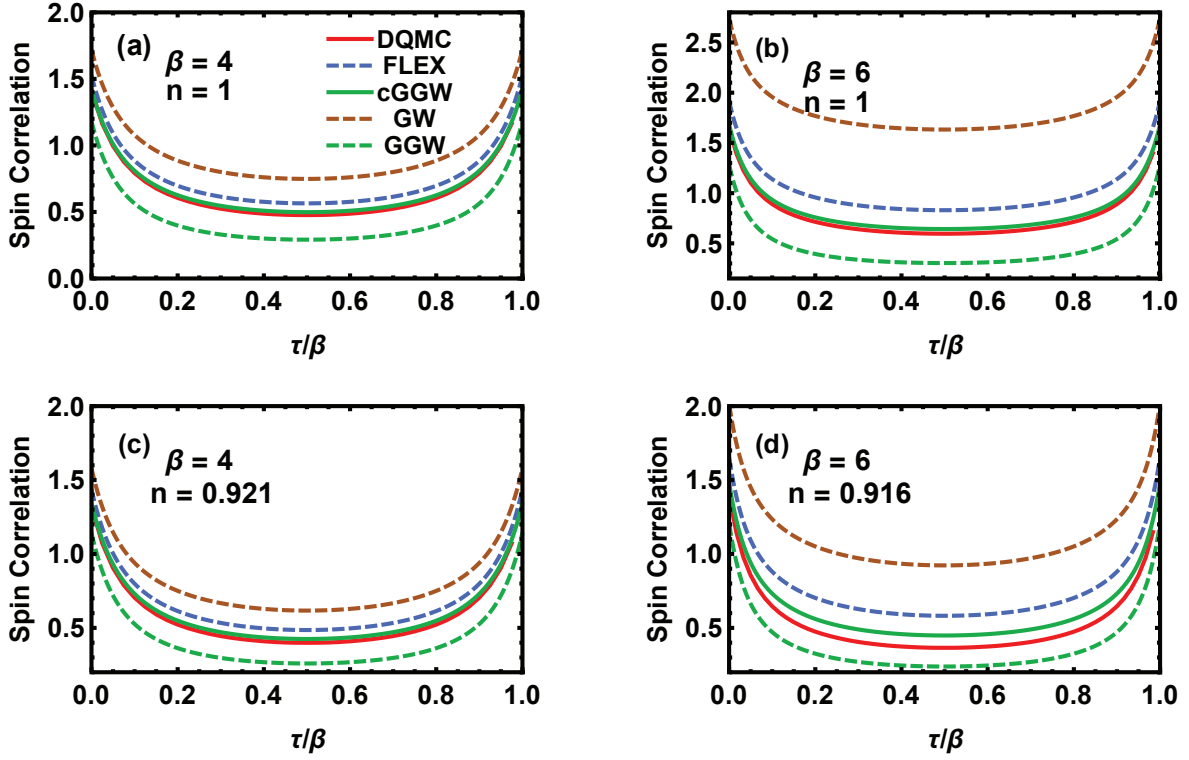}
\captionsetup{font={small},justification=raggedright}
\caption{\label{chis_tau} Antiferromagnetic spin susceptibility $\chi_{sp}(\vec k =\vec Q,\tau)$ as a function of imaginary time for DQMC, cGGW, FLEX-RPA, traditional GW-RPA and GGW-RPA at $U=2$ for different parameters: (a) $\beta=4$, $n=1$, (b) $\beta=6$, $n=1$, (c) $\beta=4$, $n=0.921$, (d) $\beta=6$, $n=0.916$. at half-filling $n=1$, corresponding to a chemical potential of $\mu=U/2=1$.
The error of DQMC is $10^{-3}$, other methods are all calculated through the discrete time algorithm with $L=16$ and $M=1024$ (almost M reaches to infinite limit). }
\end{figure}

To demonstrate the effectiveness of the method, we compare the antiferromagnetic susceptibility from the cGGW method with that obtained from the determinantal quantum Monte Carlo (DQMC) method\cite{DQMC1981,Hirsch1985,santos2003introduction}, which is numerically exact and often serves as a benchmark for approximate methods.
We consider the case of $U=2$ at half-filling and away from half-filling for different temperatures ($\beta=4$ and 6).
As shown in Fig.~\ref{chis_tau}, cGGW exhibits a high precise agreement of the imaginary time antiferromagnetic susceptibility in comparison to the DQMC benchmark, and is better than RPA results for $GW$, GGW, and fluctuation-exchange (FLEX) approximations\cite{Bickers1989}.

\begin{figure}
\includegraphics[width=0.45\textwidth]{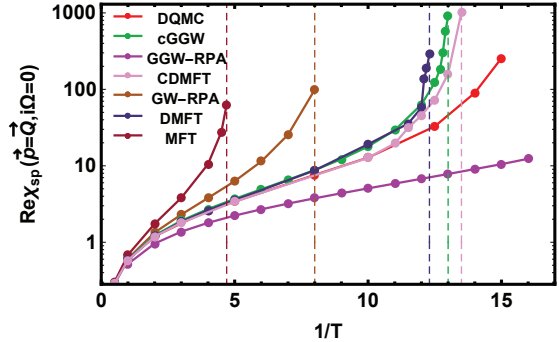}
\captionsetup{font={small},justification=raggedright}
\caption{\label{chis_T} Antiferromagnetic static susceptibility $\chi_{sp}(\vec k =\vec Q,i\omega_n=0)$ as a function of (inverse) temperature for various methods on a logarithmic scale at $U=2$ and $n=1$ in the thermodynamic limit. Data of MFT, DQMC, DMFT, CDMFT($N_c=8\times8$) is taken from \cite{PhysRevX.11.011058}.}
\end{figure}

To calculate the static antiferromagnetic susceptibility for infinite lattice, we use the finite-size scaling to approach the thermodynamic limit, and choose samples with lattice sizes from $L = 32$ to $L=128$. We take time slices M 8 values from 512 to 2048, extrapolating to infinite M results. Fig.~\ref{chis_T} shows $\chi_\mathrm{sp}(\vec Q,i\omega_n=0)$ for various methods as a function of the inverse temperatures on a logarithmic scale.
The cGGW curve (green line) displays a quantitative agreement with the numerically exact DQMC method (red line) until $\beta\approx8$. Since the thermodynamic transition manifests itself as a divergence of the susceptibility at the corresponding wave vector, our cGGW results also indicate an antiferromagnetic transition at ($\beta\approx13.1$).
Compared with other methods, the cGGW is clearly much better than the mean-field theory (MFT), $GW$-RPA, and $GGW$-RPA methods, and is comparable with the dynamical mean-field theory (DMFT) or cellular DMFT (CDMFT) methods which, however, usually require expensive computational costs.
On the contrary, the computational complexity of the cGGW susceptibility at a specific momentum is $\mathcal{O}(L^dMlog(LM))$, with $d$ the lattice dimension. For a typical parameter set in the discussion (lattice size $16\times 16$, time slices $M=1024$), the numerical cost of calculation for spin fluctuation for a single momentum and frequency is only $3.93$ seconds on a 4-core CPU(1.8GHz), indicating a computationally efficient method.

\textit{Conclusions.}---
The WTI is considered to be of particular importance, because it reflects fundamental aspects of the underlying physics, as current conservation law and gauge invariance \cite{book-peskin,Kadanoff,DMFTRMP}. For example, previously, it has been shown that the WTI plays a crucial role in obtaining the reasonable diamagnetic susceptibility \cite{Levin2018,KM1961} and optical conductivity \cite{Millis2009} of high-temperature superconductors.
In this Letter, we propose the cGGW approach to calculate the response function based on the FDT,
which also preserves the WTI automatically because the symmetry is always respected. In addition, we find that the BSE preserves the WTI but violates the FDT, whereas the RPA violates both. The preservation of WTI and  FDT in our approach are verified in the 2D Hubbard model numerically. The cGGW spin-spin correlation functions are compared with the DQMC benchmark, and the results are satisfactory.
Due to the relatively low computational cost and high effectiveness,
this method is expected to be applied in realistic material computation in the future for the calculations of the various susceptibilities, especially the transport properties of the correlated systems with spin-dependent interaction.

\begin{acknowledgments}
This work is supported by the National Natural Science Foundation of China (Grant No.12174006) and High-performance Computing Platform of Peking University. H.H. acknowledges the support of the National Key R\&D Program of China (No. 2021YFA1401600), the National Natural Science Foundation of China (Grant No. 12074006), and the start-up fund from Peking University. Authors are very grateful to B.Rosenstein, Tianxing Ma, Hong Jiang, Xinguo Ren, and Zhenhao Fan for valuable discussions and helps in numerical computations.
\end{acknowledgments}

\nocite{*}

\bibliography{cggw}

\newpage
\part*{Supplemental Material}
\setcounter{section}{0}
\setcounter{figure}{0}
\setcounter{equation}{0}
\setcounter{page}{1}
\renewcommand{\thesection}{S\arabic{section}}
\renewcommand{\thetable}{S\arabic{table}}
\renewcommand{\thefigure}{S\arabic{figure}}

\input{supple.tex}

\end{document}

%% file: supple.tex
\section{Generalized $GW$ equations in electronic systems}

The action for a general electronic system with spin-dependent action
takes the form
\begin{equation}
S\left[\psi^{\ast},\psi\right]=-\sum_{\alpha_{1}\alpha_{2}}\int d\left(12\right)\ \psi_{\alpha_{1}}^{\ast}\left(1\right)T_{\alpha_{1}\alpha_{2}}\left(1,2\right)\psi_{\alpha_{2}}\left(2\right)+\frac{1}{2}\sum_{ab}\int d\left(12\right)\ \sigma^{a}\left(1\right)V^{ab}\left(1,2\right)\sigma^{b}\left(2\right).\label{eq:action}
\end{equation}
Here $\alpha_{1},\alpha_{2}$ indicate the spin, taking values of
spin-up and spin-down. $a,b$ take values of $0,x,y,z$, and $\sigma^{a}$
is a component of the composite vector operator $\vec{\sigma}\left(1\right)\equiv\sum_{\alpha_{1}\alpha_{2}}\psi_{\alpha_{1}}^{\ast}\left(1\right)\vec{\tau}_{\alpha_{1}\alpha_{2}}\psi_{\alpha_{2}}\left(1\right)$
with $\vec{\tau}=\left(\tau^{0},\tau^{x},\tau^{y},\tau^{z}\right)$
are the Pauli's matrices. We introduce an external vector source $\vec{J}\left(1\right)$
coupled to the operator $\vec{\sigma}\left(1\right)$, and obtain
the perturbed action
\begin{equation}
S\left[\psi^{\ast},\psi;\vec{J}\right]=S\left[\psi^{\ast},\psi\right]-\sum_{a}\int d\left(1\right)\ J^{a}\left(1\right)\sigma^{a}\left(1\right).\label{eq:perturbed}
\end{equation}
 By virtue of the grand partition function $Z=\int\mathcal{D}\left[\psi^{\ast},\psi\right]\ \text{e}^{-S\left[\psi^{\ast},\psi;\vec{J}\right]}$,
one can define the one-body and two-body Green's functions as
\begin{equation}
G_{\alpha_{1}\alpha_{2}}\left(1,2\right)\equiv\left\langle \psi_{\alpha_{2}}^{\ast}\left(2\right)\psi_{\alpha_{1}}\left(1\right)\right\rangle ,\label{eq:one-body}
\end{equation}
\begin{equation}
G_{\alpha_{1}\alpha_{2}\alpha_{3}\alpha_{4}}^{\left(2\right)}\left(1,2,3,4\right)=\left\langle \psi_{\alpha_{2}}^{\ast}\left(2\right)\psi_{\alpha_{1}}\left(1\right)\psi_{\alpha_{4}}^{\ast}\left(4\right)\psi_{\alpha_{3}}\left(3\right)\right\rangle .\label{eq:two-body}
\end{equation}
Here $\left\langle \ldots\right\rangle \equiv\frac{1}{Z}\int\mathcal{D}\left[\psi^{\ast},\psi\right]\ \ldots\text{e}^{-S\left[\psi^{\ast},\psi;\vec{J}\right]}$
denotes the ensemble average. Note that
\begin{equation}
\frac{\delta G_{\alpha_{1}\alpha_{2}}\left(1,2\right)}{\delta J^{c}\left(3\right)}=\sum_{\alpha_{3}\alpha_{3}^{\prime}}G_{\alpha_{1}\alpha_{2}\alpha_{3}\alpha_{3}^{\prime}}^{\left(2\right)}\left(1,2,3,3\right)\tau_{\alpha_{3}^{\prime}\alpha_{3}}^{c}-G_{\alpha_{1}\alpha_{2}}\left(1,2\right)\sum_{\alpha_{3}\alpha_{3}^{\prime}}G_{\alpha_{3}\alpha_{3}^{\prime}}\left(3,3\right)\tau_{\alpha_{3}^{\prime}\alpha_{3}}^{c}.\label{eq:g2j}
\end{equation}

Then we address the derivation of the generalized Hedin's equations.
The equation of motion stems from the invariance of functional integral
measure under the infinitesimal translation transform of the Fermionic
fields, which leads to the equality
\begin{equation}
0=\int\mathcal{D}\left[\psi^{\ast},\psi\right]\ \frac{\delta}{\delta\psi_{\alpha_{1}}^{\ast}\left(1\right)}\left(\psi_{\alpha_{2}}^{\ast}\left(2\right)\text{e}^{-S\left[\psi^{\ast},\psi;\vec{J}\right]}\right).\label{eq:ds-equality}
\end{equation}
Then one obtains
\begin{align}
\delta\left(1,2\right)\delta_{\alpha_{1}\alpha_{2}} & =\sum_{\alpha_{3}}\int d\left(3\right)\ T_{\alpha_{1}\alpha_{3}}\left(1,3\right)G_{\alpha_{3}\alpha_{2}}\left(3,2\right)+\sum_{a\alpha_{1}^{\prime}}J^{a}\left(1\right)\tau_{\alpha_{1}\alpha_{1}^{\prime}}^{a}G_{\alpha_{1}^{\prime}\alpha_{2}}\left(1,2\right)\nonumber \\
 & \quad-\sum_{ab}\sum_{\alpha_{1}^{\prime}\alpha_{3}\alpha_{3}^{\prime}}\int d\left(3\right)\ \tau_{\alpha_{1}\alpha_{1}^{\prime}}^{a}G_{\alpha_{1}^{\prime}\alpha_{2}\alpha_{3}\alpha_{3}^{\prime}}^{\left(2\right)}\left(1,2,3,3\right)\tau_{\alpha_{3}^{\prime}\alpha_{3}}^{b}V^{ab}\left(1,3\right).\label{eq:dse-1}
\end{align}
By virtue of Eq. (\ref{eq:g2j}), one obtains
\begin{align}
\delta\left(1,2\right)\delta_{\alpha_{1}\alpha_{2}} & =\sum_{\alpha_{3}}\int d\left(3\right)\ \left(T_{\alpha_{1}\alpha_{3}}\left(1,3\right)-\Sigma_{\text{H},\alpha_{1}\alpha_{3}}\left(1,3\right)\right)G_{\alpha_{3}\alpha_{2}}\left(3,2\right)\nonumber \\
 & \quad-\sum_{ab}\sum_{\alpha_{1}^{\prime}}\int d\left(3\right)\ \tau_{\alpha_{1}\alpha_{1}^{\prime}}^{a}V^{ab}\left(1,3\right)\frac{\delta G_{\alpha_{1}^{\prime}\alpha_{2}}\left(1,2\right)}{\delta J^{b}\left(3\right)},\label{eq:dse-2}
\end{align}
with the Hartree self energy 
\begin{equation}
\Sigma_{\text{H},\alpha_{1}\alpha_{2}}\left(1,2\right)=-\delta\left(1,2\right)\sum_{a}\tau_{\alpha_{1}\alpha_{2}}^{a}v^{a}\left(1\right)\label{eq:se-h}
\end{equation}
 and the single-particle effective potential 
\begin{equation}
v^{a}\left(1\right)\equiv J^{a}\left(1\right)-\sum_{c}\int d\left(3\right)\ V^{ab}\left(1,3\right)\sum_{\alpha_{3}\alpha_{3}^{\prime}}G_{\alpha_{3}\alpha_{3}^{\prime}}\left(3,3\right)\tau_{\alpha_{3}^{\prime}\alpha_{3}}^{c}.\label{eq:v}
\end{equation}
Note that the functional derivative $\delta G/\delta J$ can be rewritten
as
\begin{equation}
\frac{\delta G_{\alpha_{1}\alpha_{2}}\left(1,2\right)}{\delta J^{c}\left(3\right)}=-\sum_{\alpha_{4}\alpha_{5}f}\int d\left(456\right)\ G_{\alpha_{1}\alpha_{4}}\left(1,4\right)\Lambda_{\alpha_{4}\alpha_{5}}^{f}\left(4,5,6\right)G_{\alpha_{5}\alpha_{2}}\left(5,2\right)\frac{\delta v^{f}\left(6\right)}{\delta J^{c}\left(3\right)},\label{eq:dg-dj}
\end{equation}
with the Hedin's vertex function
\begin{equation}
\Lambda_{\alpha_{1}\alpha_{2}}^{c}\left(1,2,3\right)\equiv\frac{\delta G_{\alpha_{1}\alpha_{2}}^{-1}\left(1,2\right)}{\delta v^{c}\left(3\right)}.\label{eq:hedin-vertex}
\end{equation}
The term $\delta v/\delta J$ 
\begin{equation}
\frac{\delta v^{a}\left(1\right)}{\delta J^{b}\left(2\right)}=\delta^{ab}\delta\left(1,2\right)-\sum_{c}\int d\left(3\right)\ V^{ac}\left(1,3\right)\sum_{\alpha_{3}\alpha_{3}^{\prime}}\left\langle \frac{\delta G_{\alpha_{3}\alpha_{3}^{\prime}}\left(3,3\right)}{\delta J^{b}\left(2\right)}\right\rangle \tau_{\alpha_{3}^{\prime}\alpha_{3}}^{c}\label{eq:dv-dj}
\end{equation}
renormalizes the two-body bare interaction to the dynamic potential:
\begin{equation}
W^{ab}\left(1,2\right)\equiv\sum_{c}\int d\left(3\right)\ \frac{\delta v^{a}\left(1\right)}{\delta J^{c}\left(3\right)}V^{bc}\left(2,3\right).\label{eq:w}
\end{equation}

For convenience, we introduce the matrix notation $\underline{X}$
for quantities with two spin indices:
\begin{equation}
\underline{X}\triangleq\left[\begin{array}{cc}
X_{\uparrow\uparrow} & X_{\uparrow\downarrow}\\
X_{\downarrow\uparrow} & X_{\downarrow\downarrow}
\end{array}\right].\label{eq:mat-nota}
\end{equation}
The one-body Green's function $G$, the Pauli matrices $\tau^{a}$
and the Hedin's vertex function $\Lambda^{a}$ are all such quantities.
With this notation and definitions (\ref{eq:hedin-vertex}, \ref{eq:w}),
Eq. (8) can be rearranged as
\begin{equation}
\underline{G}^{-1}\left(1,2\right)=\underline{T}\left(1,2\right)-\underline{\Sigma}_{\text{H}}\left(1,2\right)-\underline{\Sigma}\left(1,2\right),\label{eq:dyson}
\end{equation}
where the self energy is given by
\begin{equation}
\underline{\Sigma}\left(1,2\right)=-\sum_{ac}\int d\left(34\right)\ \underline{\tau}^{a}\underline{G}\left(1,4\right)\underline{\Lambda}^{c}\left(4,2,3\right)W^{ca}\left(3,1\right),\label{eq:se-mat}
\end{equation}
and the Hartree self energy is 
\begin{equation}
\underline{\Sigma}_{\text{H}}\left(1,2\right)=-\delta\left(1,2\right)\sum_{a}\tau^{a}v^{a}\left(1\right),\label{eq:se-h-mat}
\end{equation}
Substituting Eqs. (\ref{eq:dg-dj}, \ref{eq:dv-dj}) into Eq. (\ref{eq:w}),
one obtains
\begin{equation}
W^{ab}\left(1,2\right)=V^{ab}\left(1,2\right)+\sum_{cd}\int d\left(34\right)\ V^{ac}\left(1,3\right)P^{cd}\left(3,4\right)W^{db}\left(4,2\right),\label{eq:w-eq}
\end{equation}
where the polarization function $P$ is given by
\begin{equation}
P^{ab}\left(1,2\right)=\int d\left(34\right)\ \text{Tr}\left[\underline{\tau}^{a}\underline{G}\left(1,3\right)\underline{\Lambda}^{b}\left(3,4,2\right)\underline{G}\left(4,1\right)\right].\label{eq:p-def}
\end{equation}
Eq. (\ref{eq:w-eq}) can be rearrange in a simple form as
\begin{equation}
\left(W^{-1}\right)^{ab}\left(1,2\right)=\left(V^{-1}\right)^{ab}\left(1,2\right)-P^{ab}\left(1,2\right),\label{eq:w-eq-sim}
\end{equation}

Up to now, the generalized Hedin's equations are all derived, and
summarized as follows:
\begin{align}
\underline{G}^{-1}\left(1,2\right) & =\underline{T}\left(1,2\right)-\underline{\Sigma}_{\text{H}}\left(1,2\right)-\underline{\Sigma}\left(1,2\right),\nonumber \\
\underline{\Sigma}\left(1,2\right) & =-\sum_{ac}\int d\left(34\right)\ \underline{\tau}^{a}\underline{G}\left(1,4\right)\underline{\Lambda}^{c}\left(4,2,3\right)W^{ca}\left(3,1\right),\nonumber \\
\left(W^{-1}\right)^{ab}\left(1,2\right) & =\left(V^{-1}\right)^{ab}\left(1,2\right)-P^{ab}\left(1,2\right),\nonumber \\
P^{ab}\left(1,2\right) & =\int d\left(34\right)\ \text{Tr}\left[\underline{\tau}^{a}\underline{G}\left(1,3\right)\underline{\Lambda}^{b}\left(3,4,2\right)\underline{G}\left(4,1\right)\right].\label{eq:hedin-eqs}
\end{align}
Note that the explicit expression for the Hedin's vertex function
$\Lambda$ is unknown in Hedin's framework. 

The simplification 
\begin{equation}
\Lambda^{c}\left(1,2,3\right)\approx-\frac{\delta\Sigma_{\text{H}}\left(1,2\right)}{\delta v^{c}\left(3\right)}=\tau^{c}\delta\left(1,2\right)\delta\left(1,3\right),\label{eq:vertex-sim}
\end{equation}
yields the generalized $GW$ (GGW) approximation. In the GGW approximation,
the self energy $\Sigma$ is given by
\begin{equation}
\underline{\Sigma}\left(1,2\right)=-\sum_{ab}\underline{\tau}^{a}\underline{G}\left(1,2\right)\underline{\tau}^{b}W^{ba}\left(2,1\right),\label{eq:se-ggw}
\end{equation}
and the polarization function becomes
\begin{equation}
P^{ab}\left(1,2\right)=\mathrm{Tr}\left[\underline{\tau}^{a}\underline{G}\left(1,2\right)\underline{\tau}^{b}\underline{G}\left(2,1\right)\right].\label{eq:p-ggw}
\end{equation}

\section{Covariant $GGW$ equations}

We consider the calculation of a general two-body correlation function
$\left\langle X\left(1\right)Y\left(1\right)\right\rangle _{\text{c}}$,
where $X,Y$ are local binary operators and take the form
\begin{align}
X\left(1\right) & =\int d\left(\epsilon\epsilon^{\prime}\right)\ \sum_{a\alpha\alpha^{\prime}}x^{a}\left(\epsilon,\epsilon^{\prime}\right)\psi_{\alpha}^{\ast}\left(1+\epsilon\right)\tau_{\alpha\alpha^{\prime}}^{a}\psi_{\alpha^{\prime}}\left(1+\epsilon^{\prime}\right),\nonumber \\
Y\left(1\right) & =\int d\left(\epsilon\epsilon^{\prime}\right)\ \sum_{a\alpha\alpha^{\prime}}y^{a}\left(\epsilon,\epsilon^{\prime}\right)\psi_{\alpha}^{\ast}\left(1+\epsilon\right)\tau_{\alpha\alpha^{\prime}}^{a}\psi_{\alpha^{\prime}}\left(1+\epsilon^{\prime}\right)\text{,}\label{eq:loc_bin_op}
\end{align}
where $\epsilon$ is the deviation from coordinate $1$. The functional
derivative scheme to calculate $\left\langle X\left(1\right)Y\left(1\right)\right\rangle $
is formulated in details below.

First, add an external local source $\phi\left(1\right)$ coupled
to the operator $X\left(1\right)$ and thus the perturbed action becomes
\begin{equation}
S\left[\psi^{\ast},\psi,\phi\right]=S\left[\psi^{\ast},\psi\right]-\int d\left(1\right)\ \phi\left(1\right)X\left(1\right).\label{eq:perturb_action}
\end{equation}
The additional term $\int d\left(1\right)\ \phi\left(1\right)X\left(1\right)$
is explicitly expressed as
\begin{equation}
\int d\left(12\right)\ \sum_{a\alpha_{1}\alpha_{2}}\psi_{\alpha_{1}}^{\ast}\left(1\right)\left\{ \int d\left(\epsilon\epsilon^{\prime}\right)\ \phi\left(1-\epsilon\right)x^{a}\left(\epsilon,\epsilon^{\prime}\right)\tau_{\alpha_{1}\alpha_{2}}^{a}\delta\left(1+\epsilon^{\prime},2+\epsilon\right)\right\} \psi_{\alpha_{2}}\left(2\right).\label{eq:additional}
\end{equation}
Note that the additional term can be regarded as a variation of the
$T$ term:
\begin{equation}
\underline{T}\left(1,2;\phi\right)=\underline{T}\left(1,2\right)+\sum_{a}\int d\left(\epsilon\epsilon^{\prime}\right)\ \phi\left(1-\epsilon\right)x^{a}\left(\epsilon,\epsilon^{\prime}\right)\underline{\tau}^{a}\delta\left(1+\epsilon^{\prime},2+\epsilon\right).\label{eq:var-T}
\end{equation}

Then, calculate the functional derivative of the $GGW$ equations
with respect to the external source $\phi$. It is convenient to introduce
some vertex functions
\begin{align}
\underline{\Gamma}\left(1,2,3\right) & \equiv\frac{\delta\underline{G}^{-1}\left(1,2\right)}{\delta\phi\left(3\right)},\nonumber \\
\underline{\gamma}\left(1,2,3\right) & \equiv\frac{\delta\underline{T}\left(1,2;\phi\right)}{\delta\phi\left(3\right)},\nonumber \\
\underline{\Gamma}_{\text{H}}\left(1,2,3\right) & \equiv-\frac{\delta\Sigma_{\text{H}}\left(1,2\right)}{\delta\phi\left(3\right)},\nonumber \\
\underline{\Gamma}_{W}^{ab}\left(1,2,3\right) & \equiv\frac{\delta\left(W^{-1}\right)^{ab}\left(1,2\right)}{\delta\phi\left(3\right)}.\label{eq:def-vertex-funcs}
\end{align}
Here $\underline{\Gamma}$ is the full vertex, $\underline{\gamma}$
the bare vertex, $\underline{\Gamma}_{\text{H}}$ the ``bubble''
vertex induced from the Hartree self energy and $\underline{\Gamma}_{W}$
originated from the dynamical potential $W$. We also introduce the
notation $\dot{X}\left(1,2,3\right)\equiv\delta X\left(1,2\right)/\delta\phi\left(3\right)$.
Note that
\begin{equation}
\underline{\dot{G}}\left(1,2,3\right)=-\int d\left(45\right)\ \underline{G}\left(1,4\right)\underline{\Gamma}\left(4,5,3\right)\underline{G}\left(5,2\right),\label{eq:dot-g}
\end{equation}
and
\begin{equation}
\dot{W}^{ab}\left(1,2,3\right)=-\sum_{de}\int d\left(45\right)\ W^{ad}\left(1,4\right)\Gamma_{W}^{de}\left(4,5,3\right)W^{eb}\left(5,2\right).\label{eq:dot-w}
\end{equation}
It is not hard to derive the equations below
\begin{align}
\underline{\Gamma}\left(1,2,3\right) & =\underline{\gamma}\left(1,2,3\right)+\underline{\Gamma}_{\text{H}}\left(1,2,3\right)+\sum_{ab}\underline{\tau}^{a}\underline{\dot{G}}\left(1,2,3\right)\underline{\tau}^{b}W^{ba}\left(2,1\right)\nonumber \\
 & \quad+\sum_{ab}\underline{\tau}^{a}\underline{G}\left(1,2\right)\underline{\tau}^{b}\dot{W}^{ba}\left(2,1,3\right).\label{eq:exp-full}
\end{align}
\begin{equation}
\underline{\gamma}\left(1,2,3\right)=\sum_{a}\int d\left(\epsilon\epsilon^{\prime}\right)\ \delta\left(1-\epsilon,3\right)x^{a}\left(\epsilon,\epsilon^{\prime}\right)\underline{\tau}^{a}\delta\left(1+\epsilon^{\prime},2+\epsilon\right),\label{eq:exp-bare}
\end{equation}
\begin{equation}
\underline{\Gamma}_{\text{H}}\left(1,2,3\right)=-\delta\left(1,2\right)\sum_{ad}\int d\left(4\right)\ \underline{\tau}^{a}V^{ad}\left(1,4\right)\text{Tr}\left[\underline{\tau}^{d}\dot{G}\left(4,4,3\right)\right],\label{eq:exp-bubble}
\end{equation}
\begin{equation}
\underline{\Gamma}_{W}^{ab}\left(1,2,3\right)=-\text{Tr}\left[\underline{\tau}^{a}\underline{\dot{G}}\left(1,2,3\right)\underline{\tau}^{b}\underline{G}\left(2,1\right)\right]-\text{Tr}\left[\underline{\tau}^{a}G\left(1,2\right)\underline{\tau}^{b}\underline{\dot{G}}\left(2,1,3\right)\right].\label{eq:exp-w-vertex}
\end{equation}
By virtue of Eqs. (\ref{eq:dot-g}, \ref{eq:dot-w}), one obtains
\begin{equation}
\underline{\Gamma}\left(1,2,3\right)=\underline{\gamma}\left(1,2,3\right)+\underline{\Gamma}_{\text{H}}\left(1,2,3\right)+\underline{\Gamma}_{\text{MT}}\left(1,2,3\right)+\underline{\Gamma}_{\text{AL}1}\left(1,2,3\right)+\underline{\Gamma}_{\text{AL}2}\left(1,2,3\right),\label{eq:full-vertex}
\end{equation}
with the Maki-Thomson-like vertex
\begin{equation}
\underline{\Gamma}_{\text{MT}}\left(1,2,3\right)=-\sum_{ab}\int d\left(45\right)\ \underline{\tau}^{a}\underline{G}\left(1,4\right)\underline{\Gamma}\left(4,5,3\right)\underline{G}\left(5,2\right)\tau^{b}W^{ba}\left(2,1\right),\label{eq:mt-vertex}
\end{equation}
and two Aslamazov-Larkin-like vertex functions
\begin{align}
\underline{\Gamma}_{\text{AL}1}\left(1,2,3\right) & =-\sum_{abde}\int d\left(4567\right)\ \underline{\tau}^{a}\underline{G}\left(1,2\right)\underline{\tau}^{b}W^{bd}\left(2,4\right)W^{ea}\left(5,1\right)\nonumber \\
 & \quad\times\text{Tr}\left[\underline{\tau}^{d}\underline{G}\left(4,6\right)\underline{\Gamma}\left(6,7,3\right)\underline{G}\left(7,5\right)\underline{\tau}^{e}\underline{G}\left(5,4\right)\right],\label{eq:al-vertex-1}
\end{align}
\begin{align}
\underline{\Gamma}_{\text{AL}2}\left(1,2,3\right) & =-\sum_{abde}\int d\left(4567\right)\ \underline{\tau}^{a}\underline{G}\left(1,2\right)\underline{\tau}^{b}W^{bd}\left(2,4\right)W^{ea}\left(5,1\right)\nonumber \\
 & \quad\times\text{Tr}\left[\underline{\tau}^{e}\underline{G}\left(5,6\right)\underline{\Gamma}\left(6,7,3\right)\underline{G}\left(7,4\right)\underline{\tau}^{d}\underline{G}\left(4,5\right)\right].\label{eq:al-vertex-2}
\end{align}
Note that the ``bubble'' vertex
\begin{equation}
\underline{\Gamma}_{\text{H}}\left(1,2,3\right)=\delta\left(1,2\right)\sum_{ad}\int d\left(4\right)\ \underline{\tau}^{a}V^{ad}\left(1,4\right)\text{Tr}\left[\tau^{d}\underline{G}\left(1,5\right)\underline{\Gamma}\left(5,6,3\right)\underline{G}\left(6,2\right)\right].\label{eq:bubble-vertex}
\end{equation}

Next, let $\phi\to0$ and solve the vertex function (\ref{eq:full-vertex}).
Finally, calculate the two-body correlation function through the equation
\begin{align}
\left\langle X\left(2\right)Y\left(1\right)\right\rangle _{\text{c}} & \equiv\delta\left\langle Y\left(1\right)\right\rangle /\delta\phi\left(2\right)\nonumber \\
 & =\sum_{a}\int d\left(\epsilon\epsilon^{\prime}\right)\ y^{a}\left(\epsilon,\epsilon^{\prime}\right)\text{Tr}\left[\underline{\tau}^{a}\dot{G}\left(1+\epsilon,1+\epsilon^{\prime},2\right)\right]\nonumber \\
 & =-\int d\left(45\epsilon\epsilon^{\prime}\right)\ y\left(\epsilon,\epsilon\right)\text{Tr}\left[\underline{\tau}^{a}\underline{G}\left(1+\epsilon,4\right)\underline{\Gamma}\left(4,5,2\right)\underline{G}\left(5,1+\epsilon^{\prime}\right)\right].\label{eq:tbcf}
\end{align}

To sum up, the procedure for calculating the two-body correlation
function is as follows. First, add an external source coupled to the
specified binary operator. Then, make functional derivative of off-shell
GGW equations with respect to the source. Next, solve the on-shell
covariant GGW equations to obtain the vertex function. Finally, calculate
the two-body correlation function through Eq. (\ref{eq:tbcf}). 

\section{Implementation in the two-dimensional Hubbard model}

\subsection{Discretized Matsubara time action}

We use the discretized Matsubara time path integral formalism for
the numerical implementation. For a general normal ordered Hamiltonian
$\mathcal{H}\left[\psi^{\dagger},\psi\right]$, the discretized-time
action reads
\begin{equation}
S_{M}\left[\psi^{\ast},\psi\right]=\sum_{l=0}^{M-1}\sum_{\alpha=\uparrow,\downarrow}\sum_{\vec{x}}\left\{ \psi_{\alpha}^{\ast}\left(\vec{x},\tau_{l}\right)\left(\psi_{\alpha}\left(\vec{x},\tau_{l+1}\right)-\psi_{\alpha}\left(\vec{x},\tau_{l}\right)\right)\right\} +\sum_{l=1}^{M-1}\Delta\tau\mathcal{H}\left[\psi_{\alpha}^{\ast}\left(\vec{x},\tau_{l}\right)\psi_{\alpha}\left(\vec{x},\tau_{l}\right)\right].\label{eq:dt-action}
\end{equation}
Here $M$ is the number of time slices, and $\Delta\tau=\beta/M$.
The integer $l$ labels the discretized Matsubara time, and $\tau_{l}\equiv l\Delta\tau$. 

For the Hubbard model with the Hamiltonian
\begin{equation}
    \mathcal{H}=-t\sum_{\left< \vec{x}_{1}\vec{x}_{2}\right>\alpha}\psi_{\vec{x}_{1}\alpha}^{\dagger}\psi_{\vec{x}_{2}\alpha}\nonumber\\
    -\frac{U}{6}\sum_{\vec{x}a}\sigma_{\vec{x}}^{a}\sigma_{\vec{x}}^{a}-\left(\mu-\frac{U}{2}\right)\sum_{\alpha\vec{x}}\psi_{\alpha\vec{x}}^{\dagger}\psi_{\alpha\vec{x}},\label{eq:hub-ham}
\end{equation}
compare the action (\ref{eq:dt-action}) with the form (\ref{eq:action}),
and one obtains
\begin{equation}
\underline{T}\left(1,2\right)=\left[-\frac{1}{\Delta\tau}\delta_{\vec{x}_{1}\vec{x}_{2}}\left(\delta_{l_{1},l_{2}-1}-\delta_{l_{1},l_{2}}\right)+t_{\vec{x}_{1}\vec{x}_{2}}\delta_{l_{1},l_{2}}+\mu\delta_{\vec{x}_{1}\vec{x}_{2}}\delta_{l_{1},l_{2}}\right]\underline{\tau}^{0},\label{eq:t-form}
\end{equation}
and
\begin{equation}
V^{ab}\left(1,2\right)=\delta_{\vec{x}_{1}\vec{x}_{2}}\delta_{l_{1}l_{2}}\delta^{ab}I^{s},\label{eq:v-form}
\end{equation}
with $a,b$ taking values of $x,y,z$, and $I^{s}=-U/3$ the bare
spin potential. Here the label $1,2$ denotes for $\left(\vec{x}_{1},\tau_{1}\right),\left(\vec{x}_{2},\tau_{2}\right)$
respectively.

For a lattice with the translation symmetries, we use the discrete
Fourier transformation to simplify our equations. The Fermionic array
$X_{\text{F}}$ takes the form
\begin{equation}
X_{\text{F}}\left(1,2\right)=\frac{1}{\mathcal{N}}\sum_{k}X_{\text{F}}\left(k\right)\mathcal{E}_{\text{F}}\left(k,1-2\right),\label{eq:f-arr}
\end{equation}
and the Bosonic array $X_{\text{B}}$ takes the form
\begin{equation}
X_{\text{B}}\left(1,2\right)=\frac{1}{\mathcal{N}}\sum_{k}X_{\text{B}}\left(k\right)\mathcal{E}_{\text{B}}\left(k,1-2\right).\label{eq:b-arr}
\end{equation}
Here the transformation kernels $\mathcal{E}_{\text{F}}$ and $\mathcal{E}_{\text{B}}$
are defined as
\begin{equation}
\mathcal{E}_{\text{F}}\left(k,1-2\right)\equiv\text{e}^{\text{i}\vec{k}\cdot\left(\vec{x}_{1}-\vec{x}_{2}\right)}\text{e}^{-\text{i}\frac{2m_{k}+1}{M}\left(l_{1}-l_{2}\right)},\label{eq:f-ker}
\end{equation}
\begin{equation}
\mathcal{E}_{\text{B}}\left(k,1-2\right)\equiv\text{e}^{\text{i}\vec{k}\cdot\left(\vec{x}_{1}-\vec{x}_{2}\right)}\text{e}^{-\text{i}\frac{2m_{k}}{M}\left(l_{1}-l_{2}\right)},\label{eq:b-ker}
\end{equation}
respectively. Here $\mathcal{N}=\beta L^{2}$, $k=\left(\vec{k},m_{k}\right)$
and $m_{k}$ takes the integer value from $0$ to $M-1$. Note that
the transformation of the $T$-term is
\begin{equation}
\underline{T}\left(k\right)=\left[-\frac{1}{\Delta\tau}\left(\text{e}^{-\text{i}\pi\left(2m_{k}+1\right)/M}-1\right)-\varepsilon\left(\vec{k}\right)+\mu\right]\underline{\tau}^{0},\label{eq:t-k}
\end{equation}
with $\varepsilon\left(\vec{k}\right)$ the non-interacting dispersion.
For the two-dimensional Hubbard model, $\varepsilon\left(\vec{k}\right)=-2t\left(\cos k_{x}+\cos k_{y}\right)$
with $t$ the nearest-neighbor hopping strength. 

\subsection{GGW and covariant GGW equations in Fourier space}

Note that the one-body Green's function $\underline{G}$ and the self-energy
$\underline{\Sigma}$ are Fermionic arrays, and the dynamical potential
$W^{ab}$ and the polarization $P^{ab}$ are Bosonic arrays. It is
easy to derive the GGW equations in Fourier space
\begin{align}
\underline{G}^{-1}\left(k\right) & =\underline{T}^{-1}\left(k\right)-\underline{\Sigma}_{\text{H}}\left(k\right)-\underline{\Sigma}\left(k\right),\nonumber \\
\underline{\Sigma}\left(k\right) & =-\frac{1}{\mathcal{N}}\sum_{q,ab}\underline{\tau}^{a}\underline{G}\left(k+q\right)\underline{\tau}^{b}W^{ba}\left(q\right),\nonumber \\
\left(W^{-1}\right)^{ab}\left(q\right) & =\left(V^{-1}\right)^{ab}\left(q\right)-P^{ab}\left(q\right),\nonumber \\
P^{ab}\left(q\right) & =\frac{1}{\mathcal{N}}\sum_{k}\text{Tr}\left[\underline{\tau}^{a}\underline{G}\left(q+k\right)\underline{\tau}^{b}\underline{G}\left(q\right)\right].\label{eq:ggw-fourier}
\end{align}

To derive the covariant GGW equations in Fourier space, we first make
ansatz for the vertex function
\begin{equation}
\underline{\Gamma}\left(1,2,3\right)=\frac{1}{\mathcal{N}^{2}}\sum_{p,q}\underline{\Gamma}\left(p,q\right)\mathcal{E}_{\text{F}}\left(k,1-2\right)\mathcal{E}_{\text{B}}\left(q,1-3\right).\label{eq:ansatz-vertex}
\end{equation}
Then one obtains
\begin{equation}
\underline{\Gamma}\left(p,q\right)=\underline{\gamma}\left(p,q\right)+\underline{\Gamma}_{\text{H}}\left(p,q\right)+\underline{\Gamma}_{\text{MT}}\left(p,q\right)+\underline{\Gamma}_{\text{AL1}}\left(p,q\right)+\underline{\Gamma}_{\text{AL2}}\left(p,q\right).\label{eq:gamma-fourier}
\end{equation}
The bare vertex is
\begin{equation}
\underline{\gamma}\left(p,q\right)=\sum_{a}\int d\left(\epsilon\epsilon^{\prime}\right)\ x^{a}\left(\epsilon,\epsilon^{\prime}\right)\underline{\tau}^{a}\text{e}^{\text{i}p\cdot\left(\epsilon-\epsilon^{\prime}\right)}\text{e}^{\text{i}q\cdot\epsilon}.\label{eq:bare-vertex}
\end{equation}
The bubble vertex is
\begin{equation}
\underline{\Gamma}_{\text{H}}\left(p,q\right)=\frac{1}{\mathcal{N}}\sum_{cd}\sum_{k}\underline{\tau}^{c}V^{cd}\left(q\right)\mathrm{Tr}\left[\underline{\tau}^{d}\underline{G}\left(k+q\right)\underline{\Gamma}\left(k,q\right)\underline{G}\left(k\right)\right].\label{eq:bubble-vertex-1}
\end{equation}
The MT vertex is
\begin{equation}
\underline{\Gamma}_{\text{MT}}\left(p,q\right)=-\frac{1}{\mathcal{N}}\sum_{cd}\sum_{k}\underline{\tau}^{c}\underline{G}\left(p+k+q\right)\underline{\Gamma}\left(p+k,q\right)\underline{G}\left(p+k\right)\underline{\tau}^{d}W^{dc}\left(k\right).\label{eq:mt-vertex-1}
\end{equation}
The two AL vertices are
\begin{align}
\Gamma_{\text{AL1}}\left(k,q\right) & =-\frac{1}{\mathcal{N}^{2}}\sum_{cdef}\sum_{kk^{\prime}}\underline{\tau}^{c}\underline{G}\left(p+q+k\right)\underline{\tau}^{d}W^{de}\left(k+q\right)W^{fc}\left(k\right)\nonumber \\
 & \quad\times\mathrm{Tr}\left[\underline{\tau}^{e}\underline{G}\left(k+k^{\prime}+q\right)\underline{\Gamma}\left(k+k^{\prime},q\right)\underline{G}\left(k+k^{\prime}\right)\underline{\tau}^{f}\underline{G}\left(k^{\prime}\right)\right],\label{eq:al-vertex-1-1}
\end{align}
\begin{align}
\Gamma_{\text{AL2}}\left(k,q\right) & =-\frac{1}{\mathcal{N}^{2}}\sum_{cdef}\sum_{kk^{\prime}}\underline{\tau}^{c}\underline{G}\left(p+q+k\right)\underline{\tau}^{d}W^{de}\left(k+q\right)W^{fc}\left(k\right)\nonumber \\
 & \quad\times\text{Tr}\left[\underline{\tau}^{e}\underline{G}\left(k+q+k^{\prime}\right)\underline{\tau}^{f}\underline{G}\left(k^{\prime}+q\right)\underline{\Gamma}\left(k^{\prime},q\right)\underline{G}\left(k^{\prime}\right)\right].\label{eq:al-vertex-2-1}
\end{align}
The diagrammatics for these vertices are presented in Fig.~\ref{FDvertex}.

Note that in the random phase approximation (RPA), the RPA vertex
is given by
\begin{equation}
\underline{\Gamma}_{\text{RPA}}\left(p,q\right)=\underline{\gamma}\left(p,q\right)+\frac{1}{\mathcal{N}}\sum_{cd}\sum_{k}\underline{\tau}^{c}V^{cd}\left(q\right)\mathrm{Tr}\left[\underline{\tau}^{d}\underline{G}\left(k+q\right)\underline{\Gamma}_{\text{RPA}}\left(k,q\right)\underline{G}\left(k\right)\right].\label{eq:rpa-vertex}
\end{equation}
The RPA formula is usually used to calculate the density-density or
spin-spin correlation functions. In the Bethe-Salpeter equation approach,
the MT vertex is taken into account, but the AL vertices are neglected. 

\subsection{GGW and covariant GGW equations for the 2D Hubbard model}

For the 2D Hubbard model, $\underline{T}\left(k\right)$ takes the
form $T\left(k\right)\underline{\tau}^{0}$ and $V^{ab}\left(k\right)$
takes the form $I^{s}\delta^{ab}$ with $a,b$ taking values of $x,y,z$.
To find the paramagnetic solutions, we can make the ansatz
\begin{equation}
\underline{G}\left(k\right)=G\left(k\right)\underline{\tau}^{0},\ \underline{\Sigma}\left(k\right)=\Sigma\left(k\right)\underline{\tau}^{0},\label{eq:ansatz-1}
\end{equation}
and
\begin{equation}
W^{ab}\left(k\right)=W\left(k\right)\delta^{ab},\ P^{ab}\left(k\right)=P\left(k\right)\delta^{ab}.\label{eq:ansatz-2}
\end{equation}
The GGW equation is then simplified as
\begin{align}
G^{-1}\left(k\right) & =T\left(k\right)-\Sigma\left(k\right),\nonumber \\
\Sigma\left(k\right) & =-\frac{3}{\mathcal{N}}\sum_{q}G\left(k+q\right)W\left(q\right),\nonumber \\
W^{-1}\left(q\right) & =1/I^{s}-P\left(q\right),\nonumber \\
P\left(q\right) & = \frac{2}{\mathcal{N}}\sum_{k}G\left(p+k\right)G\left(p\right).\label{eq:ggw-ansatz}
\end{align}

The simplification of the covariant GGW equations related to the species
of correlation functions. We take the spin-spin correlation function
as an example here. The spin-spin correlation function $\chi_{\text{s}}^{ab}\left(p\right)$
relates to the vertex function through 
\begin{equation}
\chi_{\text{s}}^{ab}\left(p\right)=-\sum_{q}\text{Tr}\left[\underline{G}\left(p+q\right)\underline{\Gamma}^{a}\left(q,p\right)\underline{G}\left(q\right)\tau^{b}\right].\label{eq:chi-s}
\end{equation}
Here $\underline{\Gamma}^{a}$ refers to the vertex function corresponding
to the spin operator $\sigma^{a}$. By the ansatz $\underline{\Gamma}^{a}\left(q,p\right)=\underline{\tau}^{a}\Gamma\left(q,p\right)$, the spin-spin correlation function $\chi_{\text{s}}^{ab}\left(p\right)=-2\delta^{ab}G\left(p+q\right) \Gamma\left(p,q\right) G\left(q\right)$, and the equation for the vertex function is simplified as
\begin{equation}
\Gamma\left(p,q\right)=\gamma\left(p,q\right)+\Gamma_{\text{H}}\left(p,q\right)+\Gamma_{\text{MT}}\left(p,q\right)+\Gamma_{\text{AL1}}\left(p,q\right)+\Gamma_{\text{AL2}}\left(p,q\right),\label{eq:eq-hub-vertex}
\end{equation}
with the bare vertex $\gamma\left(p,q\right)=1$, the ``bubble''
vertex
\begin{equation}
\Gamma_{\text{H}}\left(p,q\right)=\frac{2I^{s}}{\mathcal{N}}\sum_{k}G\left(k+q\right)\Gamma\left(k,q\right)G\left(k\right),\label{eq:hub-bubble-vertex}
\end{equation}
the MT vertex
\begin{equation}
\Gamma_{\text{MT}}\left(p,q\right)=-\frac{1}{\mathcal{N}}\sum_{k}G\left(p+k+q\right)\Gamma\left(p+k,q\right)G\left(p+k\right)W\left(k\right),\label{eq:hub-mt-vertex}
\end{equation}
and two AL vertices
\begin{equation}
\Gamma_{\text{AL1}}\left(p,q\right)=\frac{2}{\mathcal{N}^{2}}\sum_{kk^{\prime}}G\left(p+q+k\right)W\left(k+q\right)G\left(k+k^{\prime}+q\right)\Gamma\left(k+k^{\prime},q\right)G\left(k+k^{\prime}\right)G\left(k^{\prime}\right)W\left(k\right),\label{eq:hub-al-vertex-1}
\end{equation}
\begin{equation}
\Gamma_{\text{AL2}}\left(p,q\right)=-\frac{2}{\mathcal{N}^{2}}\sum_{kk^{\prime}}G\left(p+q+k\right)W\left(k+q\right)G\left(k+k^{\prime}+q\right)G\left(k^{\prime}+q\right)\Gamma\left(k^{\prime},q\right)G\left(k^{\prime}\right)W\left(k\right).\label{eq:hub-al-vertex-2}
\end{equation}

Note that the calculation can be fasten by discrete Fourier transformation algorithm, and as a result, the computational complexity of the cGGW susceptibility at a specific momentum is $\mathcal{O}(L^dMlog(LM))$, with $d$ the lattice dimension.
\section{Ward-Takahashi identity for global $U\left(1\right)$ symmetry}
The invariance of the functional integral measure $D[\psi^*,\psi]$ under the infinitesimal gauge transformation of the complex field yields an equality
\begin{equation}
    \delta\int D[\psi^*,\psi]{\mathcal{F}}[\psi^*,\psi]e^{-S[\psi^*,\psi]}=0,
\end{equation}
with $\mathcal{F}$ an arbitrary functional.  

For the charge $U(1)$ rotation, one can obtain:
\begin{equation}
    \sum_{\alpha_1}\int D[\psi^*,\psi][\psi_{\alpha_1}^*(1)\frac{\delta}{\delta\psi^*_{\alpha_1}(1)}-\psi_{\alpha_1}(1)\frac{\delta}{\delta\psi_{\alpha_1}(1)}] {\mathcal{F}}[\psi^*,\psi]e^{-S}=0 .\label{WTIo}
\end{equation}
Letting $\mathcal{F}=1$ in Eq.(\ref{WTIo}) yields the WTI for the one-body Green's function:
\begin{equation}
    \int d(2)Tr[\underline{T}(1,2)\underline{G}(2,1)]=\int d(2)Tr[\underline{T}(2,1)\underline{G}(1,2)].
\end{equation}
And letting $\mathcal{F}=\psi_{\alpha_3}^*(3)\psi_{\alpha_2}(2)$ yields the WTI for the two-body Green's function:  
\begin{align}
&\sum_{\alpha_1\beta_4}\int d(4)T_{\alpha_1\beta_4}(1,4)\left< \psi^*_{\alpha_3}(3)\psi_{\alpha_2}(2)\psi^*_{\alpha_1}(1)\psi_{\beta_4}(4) \right>\nonumber\\
&-\sum_{\alpha_1\beta_4}\int d(4)T_{\beta_4\alpha_1}(4,1)\left<  \psi^*_{\alpha_3}(3)\psi_{\alpha_2}(2)\psi^*_{\beta_4}(4)\psi_{\alpha_1}(1)\right>\nonumber\\
=&-\sum_{\alpha_1}(\delta(1,3)\delta_{\alpha_1\alpha_3}-\delta(1,2)\delta_{\alpha_1\alpha_2})G_{\alpha_2\alpha_3}(2,3), \label{wti-tform}
\end{align}

For the lattice system, the $T$-term takes the form
\begin{equation}
\underline{T}(1,2)=[\delta_{\vec{x}_1,\vec{x}_2}\partial_\tau-t_{\vec x_1,\vec x_2}\delta(\tau_1,\tau_2)+\mu\delta_{\vec{x}_1,\vec{x}_2}\delta (\tau_1, \tau_2)]\underline{\tau}^0,
\end{equation}
where $t_{\vec x_1,\vec x_2}=t_{\vec x_2,\vec x_1}$ is the hoping amplitude. Suppose in Eq.~(\ref{wti-tform}), the coordinate $(4)=(1+\epsilon)$, where $(\epsilon)=(\tau_\epsilon,\vec\epsilon)$. Then Eq.~(\ref{wti-tform}) can be written as
\begin{align}
&-\mathrm{i}\sum_{\vec \epsilon}\frac{\left\langle  \psi_{\alpha_3}^*(3)\psi_{\alpha_2}(2)j_{\vec\epsilon}(1)\right\rangle}{\left|\vec \epsilon \right|}+\left<\psi^*_{\alpha_3}(3)\psi_{\alpha_2}(2) \partial_\tau\rho(1) \right>\nonumber\\
=&\sum_{\alpha_1}(\delta(1,3)\delta_{\alpha_1\alpha_3}-\delta(1,2)
\delta_{\alpha_1\alpha_2})G_{\alpha_2\alpha_3}(2,3),
\end{align}
where the the hoping contribution to the current operator \cite{smit_2002,Rosenstein} is defined by:
\begin{equation}
\vec j_{\vec\epsilon}(1)=-\mathrm{i}t_{\vec x_1,\vec x_1+\vec\epsilon}\sum_{\alpha_1}[\psi^*_{\alpha_1}(1)\psi_{\alpha_1}(1+\vec\epsilon)-\psi^*_{\alpha_1}(1+\vec\epsilon)\psi_{\alpha_1}(1)]\vec\epsilon.
\end{equation}
As an example, for a square lattice with nearest neighbor hoping $t_{\vec x_1,\vec x_2}=-t (\delta_{\vec x_1,\vec x_2+\vec e^\nu}+\delta_{\vec x_1,\vec x_2-\vec e^\nu}$, the current operator along a lattice vector $\hat e^\nu$ is :
\begin{equation}
j^\nu(1)=j_{\vec e^\nu}(1)=\mathrm{i}t\sum_{\alpha_1}[\psi^*_{\alpha_1}(1)\psi_{\alpha_1}(1+\vec e^\nu)-\psi^*_{\alpha_1}(1+\vec e^\nu)\psi_{\alpha_1}(1)]\left|\vec e^\nu\right|.
\end{equation}
One can obtain the WTI in the form of the vertex:
\begin{align}
&\text{i}\sum_{\vec\epsilon} \frac{\underline{\Gamma}_{\vec\epsilon}(1,2,3)}{\left|\vec\epsilon \right|}-\partial_{\tau_3}\underline{\Gamma}^0(1,2,3)\nonumber\\
=&\delta(3,1)\underline{G}^{-1}(3,2)-\delta(3,2)\underline{G}^{-1}(1,3).
\end{align}
where the vertex corresponding to the hoping current $j_{\vec x_\epsilon}$ is defined as:
\begin{equation}
\underline\Gamma_{\vec\epsilon;\alpha_1\alpha_2}(1,2,3)=-\int d(45)\sum_{\alpha_4\alpha_5}G^{-1}_{\alpha_1\alpha_4}(1,4)\left\langle  \psi_{\alpha_5}^*(5)\psi_{\alpha_4}(4)j_{\vec\epsilon}(3)\right\rangle G_{\alpha_5\alpha_2}^{-1}(5,2).
\end{equation}
The cGGW vertex is given in Eq.~(\ref{eq:full-vertex}). To check the WTI for the cGGW vertex, one needs the bare current vertex
\begin{equation}
\underline\gamma^\epsilon(1,2,3)=-\text{i}t_{\vec x_1,\vec x_1+\vec\epsilon}\delta(1,3)[\delta(1,2+\epsilon)-\delta(1,2-\epsilon)]\tau^0/\left|\vec\epsilon \right|,
\end{equation}
and the bare charge vertex $\underline{\gamma}^0=\delta(1,2)\delta(1,3)\tau^0$.

From the Dyson equation, 
\begin{equation}
    \begin{aligned}
        \delta(3,1)\underline{G}^{-1}(3,2)-\delta(3,2)\underline{G}^{-1}(1,3)&=[\delta(3,1)\underline{T}^{-1}(3,2)-\delta(3,2)\underline{T}^{-1}(1,3)]\\&\quad+[-\delta(3,1)\underline{\Sigma}^{-1}(3,2)+\delta(3,2)\underline{\Sigma}^{-1}(1,3)].
    \end{aligned}
\end{equation}
The first term in the right hand of side can be rewritten as:
\begin{align}
&\delta(3,1)\underline{T}^{-1}(3,2)-\delta(3,2)\underline{T}^{-1}(1,3) \nonumber\\
=& -t_{x_3,x_1}\delta(2,3)+t_{x_3,x_2}\delta(1,3)-\delta(3,1)\partial_{\tau_3}\delta(2,3)-\delta(3,2)\partial_{\tau_3}\delta(1,3)\nonumber\\
=&\text{i}\sum_{\vec\epsilon} \underline{\gamma}_{\vec\epsilon}(1,2,3)\left|\vec\epsilon \right|-\partial_{\tau_3}\underline{\gamma}^0(1,2,3)\nonumber\\
\triangleq& \Gamma^{WTI}_1
\end{align}
The second term can be rewritten as: 
\begin{align}
&-\delta(3,1)\underline{\Sigma}^{-1}(3,2)+\delta(3,2)\underline{\Sigma}^{-1}(1,3)\nonumber\\
=&-\sum_{cd}\int d(45)\underline \tau^c \underline G(1,4)[\delta(3,4)\underline G^{-1}(3,5)-\delta(3,5)\underline G^{-1}(4,3)]\underline G(5,2)\underline\tau^dW^{dc}(2,1)\nonumber\\
=&-\sum_{cd}\int d(45)\underline \tau^c \underline G(1,4)[\text{i}\sum_{\vec\epsilon} \frac{\underline{\Gamma}_{\vec\epsilon}(4,5,3)}{\left|\vec\epsilon \right|}-\partial_{\tau_3}\underline{\Gamma}^0(4,5,3)]\underline G(5,2)\underline\tau^dW^{dc}(2,1)\nonumber\\
\triangleq& \Gamma^{WTI}_2
\end{align}
We can also insert some zeros:

\begin{align}
\Gamma_3^{WTI}&=\delta(1,2)\sum_{cd}\int d(456)\underline{\tau}^cV^{cd}(1,4)\text{Tr}[\underline{\tau}^d\underline{G}(1,5)\nonumber\\
&(\delta(3,5)\underline{G}^{-1}(3,6)-\delta(3,6)\underline{G}^{-1}(5,3))\underline{G}(6,2)]\nonumber\\
&=\delta(1,2)\sum_{cd}\int d(456)\underline{\tau}^cV^{cd}(1,4)Tr[\underline{\tau}^d\underline{G}(1,5)\nonumber\\
&(\text{i}\sum_{\vec\epsilon} \frac{\underline{\Gamma}_{\vec\epsilon}(5,6,3)}{\left|\vec\epsilon \right|}-\partial_{\tau_3}\underline{\Gamma}^0(5,6,3))\underline{G}(6,2)]\nonumber\\
&=0
\end{align}

\begin{equation}
\Gamma_4^{WTI}\equiv-\sum_{cdef}\int d(567)\underline{\tau}^c \underline{G}(1,2)\underline{\tau}^dW^{de}(1,4)
\Omega W^{fc}(5,2),
\end{equation}
where
\begin{align}
\Omega&\equiv\int d(567)Tr[ \underline{\tau}^e\underline{G}(4,6)[\text{i}\sum_{\vec\epsilon} \frac{\underline{\Gamma}_{\vec\epsilon}(6,7,3)}{\left|\vec\epsilon \right|}-\partial_{\tau_3}\underline{\Gamma}^0(6,7,3)]\underline{G}(7,5)\underline{\tau}^f\underline{G}(5,4)\nonumber\\
&+ \underline{\tau}^e \underline{G}(4,5)\underline{\tau}^f\underline{G}(5,6)[\text{i}\sum_{\vec\epsilon} \frac{\underline{\Gamma}_{\vec\epsilon}(6,7,3)}{\left|\vec\epsilon \right|}-\partial_{\tau_3}\underline{\Gamma}^0(6,7,3)]\underline{G}(7,4)]\nonumber\\
&=\int d(567)Tr[ \underline{\tau}^e\underline{G}(4,6)[-\delta(3,6)\underline{G}^{-1}(3,7)+\delta(3,7)\underline{G}^{-1}(6,3)]\underline{G}(7,5)\underline{\tau}^f\underline{G}(5,4)\nonumber\\
&+ \underline{\tau}^e \underline{G}(4,5)\underline{\tau}^f\underline{G}(5,6)[-\delta(3,6)\underline{G}^{-1}(3,7)+\delta(3,7)\underline{G}^{-1}(6,3)]\underline{G}(7,4)]\nonumber\\
&=0
\end{align}
So, for the GGW Green's function, the WTI is identical to:
\begin{equation}
\text{i}\sum_{\vec\epsilon} \frac{\underline{\Gamma}_{\vec\epsilon}(1,2,3)}{\left|\vec\epsilon \right|}-\partial_{\tau_3}\underline{\Gamma}^0(1,2,3)=\Gamma^{WTI}_1+\Gamma^{WTI}_2+\Gamma^{WTI}_3+\Gamma^{WTI}_4.
\end{equation}
And one can notice the cGGW vertex and BSE vertex satisfy the WTI automatically. 

When one consider a translation invariant system,  for example, a 2D square lattice with only nearest hopping, the deviation takes 4 non-zero value, $\vec e^x,-\vec e^x,\vec e^y,-\vec e^y$. Then the WTI can be written as
\begin{equation}
\text{i}\sum_{\nu=x,y} \frac{[\underline{\Gamma}^\nu(1,2,3)-\underline{\Gamma}^\nu(1,2,3-\vec e^\nu)]}{\left|\vec e^\nu \right|}-\partial_{\tau_3}\underline{\Gamma}^0(1,2,3)
=\delta(3,1)\underline{G}^{-1}(3,2)-\delta(3,2)\underline{G}^{-1}(1,3),
\end{equation}
where the vertex along the axis $\nu=x,y$ in this system should be
\begin{equation}
\underline{\Gamma}^\nu(1,2,3)=\underline{\Gamma}_{\vec e^\nu}(1,2,3)+\underline{\Gamma}_{-\vec e^\nu}(1,2,3)
\end{equation}
and we use the relation $\underline{\Gamma}_{-\vec \epsilon}(1,2,3)=-\underline{\Gamma}_{\vec \epsilon}(1,2,3-\vec\epsilon)$ (from the definition of the current operator).
The corresponding WTI in the Fourier space is:
\begin{equation}
 \text{i}\sum_{\nu}\underline{\Gamma}^\nu(p,q)[1-e^{\text{i}q^\nu}]-\text{i}\omega_q\underline{\Gamma}^0(p,q)=\underline{G}^{-1}(p)-\underline{G}^{-1}(p+q)
\end{equation}
In the continuous limit, the WTI above takes the form:
\begin{equation}
 \sum_{\nu}q^\nu\underline{\Gamma}^\nu(p,q)-\text{i}\omega_q\underline{\Gamma}^0(p,q)=\underline{G}^{-1}(p)-\underline{G}^{-1}(p+q).
\end{equation}
It should be noted that, for the discrete time algrithm, the density operator corresponding to the WTI is defined as $\rho(\vec x_1,l_1)=\sum_{\alpha_1}\psi^*_{\alpha_1}(\vec x_1,l_1)\psi_{\alpha_1}(\vec x_1,l_1+1)$, and the bare charge vertex is:
\begin{equation}
\underline{\gamma}^0(p,q)=\underline{\tau}^0e^{-\text{i}(2m_p+1)\pi/M}.
\end{equation}
In this case, the WTI takes the form:
\begin{equation}
 \text{i}\sum_{\nu}\underline{\Gamma}^\nu(p,q)[1-e^{\text{i}q^\nu}]+\frac{e^{-\text{i}\pi2m_q/M}-1}{\Delta\tau}\underline{\Gamma}^0(p,q)=\underline{G}^{-1}(p)-\underline{G}^{-1}(p+q).
\end{equation}

\section{Numerical results}
 \begin{table}[h]
\begin{center}
\tabcolsep=70pt
\caption{\label{chiAF_T}
The first column is the inverse temperature. The second column is the static Anti-ferromagnetic spin fluctuation obtained by cGGW equations for $U=2$ at half-filling in the infinite size limit.}
\begin{tabular}{cccccccc}

 1/T&$\chi_{sp}(\vec Q,i\omega_n=0)$\\
\hline
1 & 0.60140405\\
2 & 1.22731806\\
3 & 1.91427581\\
4 & 2.71622522\\
5 & 3.69260718\\
6 & 4.92852158\\
7 & 6.56194055\\
8 & 8.80297976\\
9 & 12.2190192\\
10 & 17.8793085\\
11 & 29.2695522\\
12 & 63.4345373\\
12.5 & 122.345972\\
12.7 & 185.12082\\
12.8 & 299.384574\\
12.9 & 570.940856\\
13 & 906.204114\\
13.1 & 15463.3954\\
\end{tabular}
\end{center}   
\end{table}
In the letter, we compare the temperature dependence of the static antiferromagnetic susceptibility obtained by cGGW with results from other methods. Here we present the data after finite-size scaling in Tab.\ref{chiAF_T}.